\documentclass[11pt]{article}
\usepackage{amsmath,amssymb}
\usepackage[dvips]{graphicx}
\usepackage{bm}

\begin{document}
\begin{center}
\vspace*{3cm}
\textbf{\Large
Critical Review of Path Integral Formulation  
}

\vspace{1.5cm}
Takehisa Fujita
\footnote{e-mail: 
fffujita@phys.cst.nihon-u.ac.jp
}
\vspace{0.3cm}

Department of Physics, Faculty  of Science and Technology, Nihon University, 
Tokyo, Japan

\vspace{2cm}

\textbf{Abstract}

\end{center}

\vspace{0.1cm}

The path integral formulation in quantum mechanics corresponds to the first quantization 
since it is just to rewrite the quantum mechanical amplitude into 
many dimensional integrations over discretized coordinates $x_n$. However, the path 
integral expression cannot be connected to the dynamics of classical mechanics, 
even though, superficially, there is some similarity between them. Further, 
the field theory path integral in terms of many dimensional integrations over fields 
does not correspond to the field quantization. We clarify the essential difference between 
Feynman's original formulation of path integral in QED and the modern version of the path 
integral method prevailing in lattice field theory calculations, and show that the former 
can make a correct second quantization while the latter cannot quantize fields at all 
and its physical meaning is unknown.


\section{Introduction}

The path integral expression in quantum mechanics   
can be obtained by rewriting the quantum mechanical amplitude 
$ K(x,x':t)= \langle x'|e^{-iHt} |  x \rangle $  
into many dimensional integrations over discretized coordinates $x_n$. 
This formulation can be related to physical observables and 
we discuss a good example of harmonic oscillator which can be evaluated analytically.  
However, one should be careful for the statement that the path integral formulation 
can be connected to the dynamics of classical mechanics. At a glance, one may feel 
that the path integral formulation can be written in terms of the Lagrangian 
of the classical mechanics. However, one can easily convince oneself that 
the path integral expression cannot be related to the time derivative of 
the two coordinates $x_k$ and $x_{k-1}$ since they have to be varied independently 
from $-\infty$ to $\infty$ in the path integral formulation. 
On the other hand, if it is a time derivative, then the difference of $(x_k -x_{k-1})$ 
should be kept sufficiently small within a distance $\Delta t$ in this context. 
The interesting point of the path integral is that, if one starts from the Lagrangian 
in classical mechanics, then one can quantize the system by the path integral 
formulation in which one does not have to solve the Schr\"odinger equation, 
but it has nothing to do with the dynamics of classical mechanics such as 
the summation of classical paths. 

Then, we show Feynman's formulation of the field theory path integral in 
quantum electrodynamics (QED), which is based on 
many dimensional integrations over the parameters  $q_{\bm{k},\lambda}$ appearing 
in the vector potential. This should be indeed connected  to the second quantization 
in field theory models since the procedure is essentially based on the path integral 
formulation in quantum mechanics of parameter space. 
However, the field theory path integral formulation in most of the textbooks is normally 
defined in terms of many dimensional integrations over fields. 
In this case, the path integral method does not correspond to the field 
quantization. Here, we clarify what should be the problems of the path integral 
formulation over the field variables and why the integrations over fields do not 
correspond to the field quantization. 

\vspace{1cm}
\section{Path Integral in Quantum Mechanics}
The path integral formulation in one dimensional quantum mechanics 
starts from the amplitude $ K(x,x':t) $  which is defined by
$$ K(x,x':t) = \langle x'|e^{-iHt} |  x \rangle \eqno{(2.1)} $$
where the system is specified by the Hamiltonian $H$. 
In the field theory textbooks, one often finds the expression of the amplitude 
$ K(x,x':t) $ in terms of the transition 
between the state $ |x, t \rangle $ and $ |x', t' \rangle $ as
$$    \langle x',t' |x, t \rangle \rightarrow \langle x'|e^{-iH(t'-t)} |  x \rangle. 
\eqno{(2.2)} $$
However, the state $ |x, t \rangle $ is not an eigenstate of the Hamiltonian and 
therefore, one cannot prove the rightarrow of eq.(2.2).  
Instead, we should rewrite the amplitude $ K(x,x':t) $  so as to understand 
its physical meaning  
$$ K(x,x':t) = \langle x'|e^{-iHt} |  x \rangle 
=\sum_n \psi_n(x') \psi_n^\dagger(x) e^{-iE_n t} \eqno{(2.3)} $$
where $\psi_n(x)$ and $E_n$ should be the eigenstate and the eigenvalue 
of the Hamiltonian $H$. We note that eq.(2.3) is not yet directly related to physical 
observables. 

\subsection{Path Integral Expression}

We start from the amplitude $ K(x',x:t)$ 
$$ K(x',x:t)=\langle x' | e^{-iHt} | x \rangle    $$
where the Hamiltonian in one dimension is given as
$$ H={ \hat{p}^2\over 2m} +U(x)  =-{1\over 2m}{\partial^2\over{\partial x^2}} +U(x).
  \eqno{(2.4)} $$
Here, a particle with its mass $m$ is bound in the potential $U(x)$. 
Now, one can make $n$ partitions of $t$ and $x'-x$, and therefore 
we label the discretized coordinate $x$ as 
$$ x=x_0,\  x_1,\  x_2,\  \cdots, \ x'=x_n . \eqno{(2.5)}  $$ 
In this case, we assume that each $x_i$ and $p$ should satisfy 
the following completeness relations
$$ \int_{-\infty}^\infty dx_i | x_{i} \rangle \langle x_{i} | =1,  \ 
 \langle x_i | p \rangle = {1\over \sqrt{2\pi}}e^{i p x_i }, 
 \  \int_{-\infty}^\infty dp | p \rangle \langle p | =1, 
 \ (i=1,\cdots, n). \eqno{(2.6)}  $$
Therefore, $ K(x',x:t)$ becomes 
$$  K(x',x:t) =\int_{-\infty}^\infty dx_1 \cdots \int_{-\infty}^\infty dx_{n-1}\times $$ 
$$ \langle x' | e^{-iH \Delta t} 
| x_{n-1} \rangle \langle x_{n-1} | e^{-iH \Delta t} | x_{n-2} \rangle \cdots 
\langle x_{1} | e^{-iH \Delta t} | x \rangle  \eqno{(2.7)}  $$
where $\Delta t$ is defined as ${ \Delta t ={t\over n} }$. 
Further, one can calculate the matrix elements, for example, as
$$ \langle x_{1} | e^{-iH \Delta t} | x \rangle = 
\langle x_{1} |  \exp \left[ -i\left(-{\Delta t \over 2m}{\partial^2\over{\partial x^2}} 
 +U(x) \Delta t \right)  \right] | x \rangle $$
$$ \simeq \exp(-iU(x) \Delta t ) \langle x_{1} |  
\exp \left( i{\Delta t \over 2m}{\partial^2\over{\partial x^2}}   \right) 
 | x \rangle +O( (\Delta t)^2 ).  \eqno{(2.8)}   $$
In addition, $\displaystyle{  \langle x_{1} |  e^{ i{\Delta t \over 2m}{\partial^2
\over{\partial x^2}}   }  | x \rangle }$ can be evaluated by inserting a complete set 
of momentum states
$$ \langle x_{1} |  \exp \left( i{\Delta t \over 2m}{\partial^2\over{\partial x^2}}   
\right)   | x \rangle = \int_{-\infty}^\infty 
{dp\over 2\pi} e^{ -i{p^2\over 2m}  \Delta t} e^{-ip(x-x_1) }  = 
\sqrt{ m\over 2i\pi\Delta t} e^{-i{m(x-x_1)^2\over{ 2\Delta t} }}.  
\eqno{(2.9)}  $$
Therefore, one finds now the path integral expression for $ K(x',x:t)$
$$   K(x',x:t) =   \lim_{n\rightarrow \infty} \left( {m\over 2i\pi\Delta t } 
\right)^{n\over 2} \times $$
$$\int_{-\infty}^\infty dx_1 \cdots \int_{-\infty}^\infty dx_{n-1} 
\exp \left\{ i\sum_{k=1}^n \left({m(x_k-x_{k-1})^2\over{ 2\Delta t} }
 -U(x_k)\Delta t\right)  \right\}  \eqno{(2.10a)}  $$
where $x_0=x$ and $x_n=x'$, respectively. Since the classical action $S$ is given as 
$$ S=\int_0^t dt \left( {1\over 2}m {\dot x}^2 -U(x) \right) 
= \lim_{n\rightarrow \infty} \sum_{k=1}^n \Delta t \left\{ 
 {m\over 2}  \left( {x_k-x_{k-1}\over \Delta t}\right)^2 
 -U(x_k) \right\}  \eqno{(2.11)} $$
the amplitude can be symbolically written as
$$   K(x',x:t) = {\cal N}\int [{\cal D}x] \exp \left\{ i\int_0^t  
\left( {1\over 2}m {\dot x}^2 -U(x) \right) dt' \right\}  \eqno{(2.10b)}  $$
where ${\cal N}\int [{\cal D}x] $ is defined as
$$ {\cal N}\int [{\cal D}x] \equiv \lim_{n\rightarrow \infty} 
\left( {m\over 2i\pi\Delta t } \right)^{n\over 2} 
\int_{-\infty}^\infty dx_1 \cdots \int_{-\infty}^\infty dx_{n-1}.  $$
This is indeed amazing in that the quantum mechanical amplitude seems to be 
connected to the Lagrangian of the classical mechanics for a particle with 
its mass $m$ in the same potential 
$U(x)$. Since the procedure of obtaining eq.(2.10a) is just to rewrite the 
amplitude by inserting the complete set of the $| x_{n} \rangle $ states, 
there is no mathematical problem involved in evaluating eq.(2.10a). 

\subsection{Physical meaning of path integral}
However, the physical meaning of the result and the procedure in obtaining eqs.(2.10) 
is not at all easy to understand. 
It is clear that eq.(2.10a) is well defined and there is no problem 
since it simply involves mathematics. However, there is a big jump from  eq.(2.10a) 
to  eq.(2.10b), even though it looks straightforward. Eq.(2.10b) indicates 
that the first term of eq.(2.10b) in the curly bracket is the kinetic energy of 
the particle in classical mechanics. In this case, however, $x_k$ and $x_{k-1}$ 
cannot be varied independently as one sees it from classical mechanics since 
it is related to the time derivative. On the other hand, they must be varied 
independently in the original version of eq.(2.10a) since it has nothing to do 
with the time derivative in the process of the evaluation. This is clear since, 
in quantum mechanics, time and coordinate are independent from each other. 
Therefore, it is difficult to interpret the first term of eq.(2.10a) as the kinetic 
energy term in classical mechanics. 
Secondly, the procedure of rewriting the amplitude is closely 
connected to the fact that the kinetic energy of the Hamiltonian is quadratic 
in $p$, that is, it is described as ${  {p^2\over 2m}}$. 
In this respect, the fundamental ingredients of the path integral formulation 
must lie in eq.(2.9) which relates the momentum operator $p^2$ to the time derivative 
of the coordinate $x$ under the condition that $x_k$ and $x_{k-1}$ are sufficiently 
close to each other. In this sense, if the kinetic energy operator were linear in $p$ 
like the Dirac equation, then there is no chance to rewrite the amplitude since 
the Gaussian integral is crucial in evaluating the integral. 

Therefore, it should be difficult to claim that eq.(2.10a) can correspond to 
the dynamics of classical mechanics, even though, superficially, there is some 
similarity between them. In other words, it is hard to prove that the quantum 
mechanical expression of  $ K(x',x:t) $ is related to any dynamics of classical mechanics. 
One may say that  $ K(x',x:t) $ happens to have a similar shape to classical Lagrangian, 
mathematically, but, physically it has nothing to do with the dynamics of 
classical mechanics. 
 
\subsubsection{No summation of classical path}
In some of the path integral textbooks, one finds the interpretation that 
the quantum mechanical dynamics can be obtained by summing up all possible paths 
in the classical mechanical trajectories. However, if one starts from eq.(2.10b) and 
tries to sum up all the possible paths, then one has to find the functional dependence 
of the coordinates on time and should integrate over all the possible coordinate 
configurations as the function of time. This should be quite different from the expression 
of eq.(2.10a). If one wishes to sum up all the possible paths in eq.(2.10b), then one may 
have to first consider the following expression of the coordinate $x$ as the function of 
time
$$ x= x_{cl}(t') +\sum_{n=1}^\infty y_n \sin \left({2\pi n\over t}\right) t' $$
where $ x_{cl}(t') $ denotes the classical coordinate that satisfies the Newton 
equation of motion with the initial conditions of $x_{cl}(0)=x$ and 
$x_{cl}(t)=x'$. The amplitude $y_n$ is the expansion coefficient. 
Therefore, the integrations over all the paths should mean that one should integrate over 
$$  \prod_{n=1}^\infty \int_{-\infty}^{\infty} d y_n $$ 
and, in this case, one can easily check that the calculated result of 
the integration over all the paths cannot reproduce the proper quantum mechanical 
result for the harmonic oscillator case. This simply means that the integration over 
$ dx_1 \cdots dx_{n-1}  $ in eq.(2.10a) and the integration over classical paths are 
completely different from each other, which is a natural result. 
This fact is certainly known to some careful physicists, but most of 
the path integral textbooks are reluctant to putting emphasis on the fact 
that the classical trajectories should not be summed up in the path integral 
formulation.  Rather, they say that the summation of all the classical paths 
should correspond to the quantization by the path integral, which is a wrong and 
misleading statement. 

\subsection{Advantage of path integral}
What should be any merits of the path integral formulation ? 
Eqs.(2.10) indicate that one can carry out the first quantization of 
the classical system once 
the Lagrangian is given where the kinetic energy term should have a 
quadratic shape, that is, $c \dot{x}^2 $ with $c$ some constant. 
In this case, one can obtain the quantized expression just by tracing back 
from eq.(2.10b) to eq.(2.10a). 

There is an advantage of the path integral formulation. That is, one does not have 
to solve the differential equation. Instead, one should carry out many dimensional 
integrations. It is, of course, not at all clear whether the many dimensional 
integrations may have some advantage over solving the differential equation or not. 
However, one can, at least, claim that the procedure of the many dimensional 
integrations is indeed an alternative method to solving the Schr\"odinger equation.

\subsection{Harmonic Oscillator Case}

When the potential $U(x)$ is a harmonic oscillator 
$$   U(x)={1\over 2}m\omega^2 x^2   $$ 
then one can evaluate the amplitude analytically after some 
lengthy calculations 
$$  K(x',x:t) = \sqrt{ m\omega\over 2i\pi \sin \omega t} 
\exp \left\{i{m\omega\over{ 2} }\left[ ({x'}^2+x^2 )\cot \omega t -
{2x'x\over \sin \omega t} \right] \right\}.  \eqno{(2.12)}  $$
On the other hand, one finds 
$$ K(x,x:t)=\langle x | e^{-iHt} | x \rangle =\sum_n e^{-iE_n t} 
|\psi_n (x)|^2 . \eqno{(2.13)}  $$
Therefore, if one integrates $ K(x,x:t) $ over all space, then one obtains
$$ \int_{-\infty}^\infty dx K(x,x:t) =\sum_{n=0}^\infty e^{-iE_n t} = 
\int_{-\infty}^\infty dx \sqrt{ m\omega\over 2i\pi \sin \omega t} 
e^{-im\omega x^2 \tan {\omega t \over 2}}={1\over 2i\sin  {\omega t \over 2} } .
 \eqno{(2.14)}  $$
Since the last term can be expanded as 
$$ {1\over 2i\sin  {\omega t \over 2} } = {e^{-{i\over 2}\omega t}
\over{1-e^{-i\omega t}}} =e^{-{i\over 2}\omega t}\sum_{n=0}^\infty (e^{-i\omega t} )^n 
= \sum_{n=0}^\infty e^{-i{\omega }t 
(n+{1\over 2} ) } \eqno{(2.15)}   $$
one obtains by comparing two equations 
$$ E_n=\omega \left(n+{1\over 2} \right) \eqno{(2.16)}  $$
which is just the right energy eigenvalue of the harmonic oscillator potential 
in quantum mechanics. 

It should be important to note that the evaluation of eq.(2.12) is entirely based on 
the expression of eq.(2.10a) which is just the quantum mechanical equation. 
Therefore, this example of the harmonic oscillator case shows that the rewriting 
of the amplitude $  K(x',x:t) $ is properly done in obtaining eq.(2.10a). 
This does not prove any connection of the $  K(x',x:t) $ to the classical 
mechanics. 

\vspace{1cm}

\section{Path integral in field theory}

The basic notion of the path integral was introduced by Feynman \cite{feyn1,feyn2,feyn3}, 
and the formulation 
of the path integral in quantum mechanics is given in terms of many dimensional 
integrations of the discretized coordinates $x_n$. 
As one sees from eq.(2.10), the amplitude is expressed in terms of many 
dimensional integrations with the weight factor of $ e^{iS}$ where $S$ is the action of 
the classical mechanics. This was, of course, surprising and interesting. However, as 
Feynman noted in his original papers, the path integral expression is not more than 
the ordinary quantum mechanics. 

When the classical particle interacts with the electromagnetic field $\bm{A}$, 
the amplitude of the particle can be expressed in terms of the many 
dimensional integrations of the action of the classical particle. However, 
the electromagnetic field $\bm{A}$ is already a quantum mechanical object, and  
therefore, there is no need for the first quantization in the Maxwell equation.  
However, when one wishes to treat physical processes which involve the absorption 
or emission of photon, then one has to quantize the  electromagnetic field 
$\bm{A}$ which is called field quantization or second quantization. 

\subsection{Field quantization }

The field quantization of the electromagnetic field $\bm{A}$ in the standard procedure 
in field theory can be done by expanding the field $\bm{A}$ in terms of the plane wave 
solutions 
$$ {\bm{A}}(x)=\sum_{\bm{k}} \sum_{\lambda =1}^2{1\over{\sqrt{2V\omega_{\bm{k}}}}}
\bm{\epsilon}(\bm{k},\lambda) \left[ c_{\bm{k},\lambda} e^{-ikx} +
  c^{\dagger}_{\bm{k},\lambda} e^{ikx} \right].  \eqno{(3.1)} $$
The field quantization requires that 
$ c_{\bm{k},\lambda}$ and  $c^{\dagger}_{\bm{k},\lambda}$ should be operators 
which satisfy the following commutation relations
$$ [ c_{\bm{k},\lambda}, \  c_{\bm{k}',\lambda'}^\dagger  ] = \delta_{ \bm{k}, \bm{k}' }
\delta_{\lambda, \lambda'} \eqno{(3.2)}  $$
and all other commutation relations should vanish. 
This is the standard way of the second quantization procedure even though 
it is not understood well from the fundamental principle. However, it is 
obviously required from experiments since electron emits photon when 
it decays from the $2p_{1\over 2}$ state to  the $1s_{1\over 2}$ state 
in hydrogen atom. 

\subsection{Field quantization in path integral (Feynman's ansatz)}

In his original paper, Feynman proposed a new method to quantize the 
electromagnetic field $\bm{A}$ in terms of the path integral formulation 
\cite{feyn1,feyn2,feyn3}. 
Here, we should first describe his formulation of the path integral. 
For the fermion part, he employed the particle expression, and 
therefore the path integral is defined in terms of quantum mechanics. 

For the gauge field, Feynman started from the Hamiltonian formulation of 
the electromagnetic field. The Hamiltonian of the electromagnetic field can be 
expressed in terms of the sum of the harmonic oscillators
$$ H_{el} ={1\over 2} \sum_{\bm{k},\lambda} \left( p_{\bm{k},\lambda}^2
+k^2 q_{\bm{k},\lambda}^2 \right) \eqno{(3.3)}  $$
where $p_{\bm{k},\lambda}$ is a conjugate momentum to $q_{\bm{k},\lambda}$. 
Here, it should be noted that  the $q_{\bm{k},\lambda}$ corresponds 
to the amplitude of the vector potential $ {\bm{A}}(x)$. The classical 
$ c_{\bm{k},\lambda}$ and  $c^{\dagger}_{\bm{k},\lambda}$ can be expressed 
in terms of  $p_{\bm{k},\lambda}$ and $q_{\bm{k},\lambda}$ as
$$ c_{\bm{k},\lambda} ={1\over{\sqrt{ 2 \omega_{\bm{k}} }}}\left( 
p_{\bm{k},\lambda}-i\omega_{\bm{k}}  q_{\bm{k},\lambda} \right)  \eqno{(3.4a)}  $$
$$ c_{\bm{k},\lambda}^\dagger ={1\over{\sqrt{ 2 \omega_{\bm{k}} }}}\left( 
p_{\bm{k},\lambda}+i\omega_{\bm{k}}  q_{\bm{k},\lambda} \right).  \eqno{(3.4b)}  $$
In this case, the Hamiltonian can be written in terms of 
$ c_{\bm{k},\lambda}$ and  $c^{\dagger}_{\bm{k},\lambda}$ as
$$ H_{el} ={1\over 2} \sum_{\bm{k},\lambda} \omega_{\bm{k}} 
\left( c^{\dagger}_{\bm{k},\lambda}c_{\bm{k},\lambda} +c_{\bm{k},\lambda}
c^{\dagger}_{\bm{k},\lambda} \right). \eqno{(3.5)}  $$
It should be important to note that the Hamiltonian of eq.(3.3) is originated 
from the Hamiltonian of field theory and it has nothing to do with the classical 
Hamiltonian of Newton dynamics. For the  electromagnetic field,  
there is no corresponding Hamiltonian of the Newton dynamics. 

\subsubsection{Feynman's ansatz}

Feynman proposed a unique way of carrying out the field quantization 
\cite{feyn1,feyn2,feyn3}. 
Since the Hamiltonian of the electromagnetic field can be written as the sum of 
the harmonic oscillators, he presented the path integral formulation for 
the electromagnetic field in terms of the Lagrangian in parameter space
$$ K(q_{\bm{k},\lambda},q_{\bm{k},\lambda}',t) \equiv
 {\cal N}\int [{\cal D}q_{\bm{k},\lambda}] \exp \left\{    
{i\over 2}\int_0^t \sum_{\bm{k},\lambda} \left( {\dot q}_{\bm{k},\lambda}^2
-k^2 q_{\bm{k},\lambda}^2 \right)  dt \right\}  \eqno{(3.6)}  $$
which should correspond to the quantization of the variables 
$q_{\bm{k},\lambda}$, and this corresponds to the quantization of 
the $ c_{\bm{k},\lambda}$ and  $c^{\dagger}_{\bm{k},\lambda}$. 
Thus, it is the second quantization of the electromagnetic field. 
As one sees from the above equation, the quantization procedure in the path 
integral formulation has nothing to do with the dynamics of classical mechanics, 
and this is most clear in eq.(3.6) since there is no corresponding classical 
mechanics in Maxwell equations.  

In this expression, there is an important assumption for the coordinates 
$q_{\bm{k},\lambda}$ which are the parameters appearing in the vector potential. 
That is, the states $ |  q_{\bm{k},\lambda} \rangle $ should make a complete set. 
Only under this assumption, one can derive the quantization of the harmonic 
oscillators.  Even though this is the parameter space, we believe that 
the assumption of the completeness of the states  $ |  q_{\bm{k},\lambda} \rangle $ 
should be reasonable. In this way, Feynman made use of the path integral expression 
to obtain the Feynman rules in the perturbation theory for QED. 
It may also be important to note that the path integral in Feynman's method 
has nothing to do with the integration of the configuration space. It is clear 
that one should not integrate out over the configuration space 
in the path integral since  the field quantization should be done 
for the parameters $ c_{\bm{k},\lambda}$ and  $c^{\dagger}_{\bm{k},\lambda}$. 

\subsection{Electrons interacting through gauge fields}

When one treats the system in which electrons are interacting through 
electromagnetic fields, one can write the whole system in terms of 
the path integral formulation. In this case, however, we treat electrons 
in the non-relativistic quantum mechanics. The electromagnetic fields are 
treated just in the same way as the previous section. 
$$ K(q_{\bm{k},\lambda},q_{\bm{k},\lambda}',\bm{r},\bm{r}',t) \equiv
 {\cal N} \int [{\cal D} \bm{r}]  [{\cal D}q_{\bm{k},\lambda}]  \times $$
$$ \exp \left\{ i\int_0^t  \left( {1\over 2}m  \dot{\bm{r}}^2 -g  \dot{\bm{ r}} 
\cdot \bm{A}(\bm{r})  
+ {1\over 2}  \sum_{\bm{k},\lambda} \left( {\dot q}_{\bm{k},\lambda}^2
-k^2 q_{\bm{k},\lambda}^2 \right)\right)  dt \right\}  \eqno{(3.7)}  $$
where the vector potential  $\bm{A}$ is given in eq.(3.1) 
$$ {\bm{A}}(x)=\sum_{\bm{k}} \sum_{\lambda =1}^2 
{\bm{\epsilon}(\bm{k},\lambda)\over\sqrt{V}  \omega_{\bm{k}} } 
 \left[ \dot{q}_{\bm{k},\lambda} \cos(\bm{k}\cdot \bm{r})
 + \omega_{\bm{k}}  q_{\bm{k},\lambda} \sin(\bm{k}\cdot \bm{r}) \right]   
\eqno{(3.8)} $$
which is now rewritten in terms of the variables $q_{\bm{k},\lambda}$. 

It should be noted that the path integral formulation works only for
the electromagnetic field since its field Hamiltonian can be described by the sum 
of the harmonic oscillators. This is a very special case, and there is only little 
chance that one can extend his path integral formulation to other field theory models. 
In particular, it should be hopeless to extend the path integral formulation 
to the field quantization of quantum chromodynamics (QCD) since QCD includes fourth 
powers of $q_{\bm{k},\lambda}$. In this case, one cannot carry out the Gaussian integral 
in the parameter space of $q_{\bm{k},\lambda}$.

\vspace{1cm}
\section{Problems in field theory path integral}

In this section, we discuss the problems in the standard treatment of the path 
integral formulation in field theory models in most of the path integral textbooks. 
Normally, one starts from writing the path integral formulation in terms of the many 
dimensional integrations over field variables. 

\subsection{Real scalar field as an example}

For simplicity, we take a real scalar field in 1+1 dimensions. 
In most of the field theory textbooks, the amplitude $Z$ is written as
$$ Z = {\cal N} \int [{\cal D} \phi(t,x)] \exp \left[ i\int 
{\cal L}(\phi, \partial_\mu \phi) dtdx \right]
\eqno{(4.1)} $$
where the Lagrangian density ${\cal L}(\phi, \partial_\mu \phi) $ is given as 
$$ {\cal L}(\phi, \partial_\mu \phi) ={1\over 2}\left({\partial \phi\over \partial t}
\right)^2 -{1\over 2} \left({\partial \phi\over \partial x}\right)^2 
-{1\over 2} m^2 \phi^2 . \eqno{(4.2)} $$
If we rewrite the path integral definition explicitly in terms of the field variable 
integrations like eq.(2.10), we find
$$  Z= {\cal N} \lim_{n\rightarrow \infty} \int_{-\infty}^\infty \cdots 
\int_{-\infty}^\infty   \prod_{k,\ell =1}^n d\phi_{k,\ell}\times   $$
$$ \exp \left[  i \sum_{k,\ell=1}^n \Delta t\Delta x 
\left( {(\phi_{k,\ell}-\phi_{k-1,\ell})^2 \over{ 2 (\Delta t)^2} } 
- {(\phi_{k,\ell}-\phi_{k,\ell -1})^2 \over{ 2(\Delta x)^2} } 
-{1\over 2}m^2 \phi_{k,\ell}^2   
\right) \right] \eqno{(4.3)}  $$
where $\phi_{k,\ell}$ is defined as
$$ \phi_{k,\ell} =\phi(t_k, x_\ell ), \ \ {\rm with} \ \  
t_1=t, \cdots ,t_n=t' \ \ {\rm and} \ \  x_1=x, \cdots ,x_n=x' . \eqno{(4.4)} $$
Also, $\Delta t$ and $\Delta x$ are defined as
$$ \Delta t ={(t-t')\over n}, \ \ \ \Delta x ={(x-x')\over n} . \eqno{(4.5)} $$
Now, we should examine the physical meaning of the expression of the amplitude $Z$ 
in eq.(4.3), and clarify as to what are the problems in eq.(4.3) 
in connection to the field quantization. The first problem is that eq.(4.3) 
does not contain any quantity which is connected to the initial and final states.  
This is clear since, in eq.(4.3), one should integrate over fields as defined and 
therefore no information of $\phi (t,x)$ and $\phi (t',x')$ is 
left while, in quantum mechanics version, the amplitude is described by the quantities 
$\psi_n(x)$ and ${\psi}^\dagger_n(x')$ which are the eigenstates of the Hamiltonian. 
The second problem is that the calculated result of the 
amplitude $Z$ must be only a function of $m, \Delta t, \Delta x $, that is
$$ Z=f(m,\Delta t, \Delta x).  \eqno{(4.6)} $$
This shows that the formulation which is started from many dimensional integrations 
over the field $\phi(t,x)$ has nothing to do with the second quantization. In addition, 
$Z$ depends on the artificial parameters $\Delta t$ and $\Delta x $, and this clearly 
shows that it cannot be related to any physical observables. 

This is in contrast with the formulation of eq.(3.7) where the  amplitude $K$ is 
specified by the quantum number of the parameter space $q_{\bm{k},\lambda}$ which is 
connected to the state with a proper number of photons, and it must be a function of 
$m,g,q_{\bm{k},\lambda}$
$$ K     =f(m,g,q_{\bm{k},\lambda},q_{\bm{k},\lambda}').   \eqno{(4.7)} $$
In addition, the  $K$ does not depend on the parameters $ \Delta t$ and $\Delta x $, 
which is a natural result as one can see it from the quantum mechanical path integral 
formulation. 

Finally, we note that the treatment of Feynman is based on the total QED 
Hamiltonian which is a conserved quantity. On the other hand, eq.(4.3) is based on 
the action which is obtained by integrating the Lagrangian density over space and time. 
As one knows, the Lagrangian density is not directly related to physical observables. 
Therefore, unless one can confirm that the path integral of field theory is reduced 
to the quantum mechanical amplitude like eq.(2.10), 
one cannot make use of the field theory path integral formulation. 
In fact, one cannot rewrite the expression of eq.(4.3) in terms of the field 
theory Hamiltonian density $\cal H$, contrary to the path integral 
formulation in quantum mechanics since one has to prepare a corresponding Fock space 
in the quantum field theory model.  
This shows that the amplitude $Z$ has nothing to do with the 
amplitude $K$ in eq.(3.7). In this respect, the amplitude $Z$ 
has no physical meaning, and therefore one can not calculate any physical quantities from 
the path integral formulation of $Z$ in field theory.

\subsection{Lattice field theory}

Most of the numerical calculations in the lattice field theory are 
based on the path integral formulation of eq.(4.3) \cite{wilson}. Unfortunately, 
the path integral formulation of eq.(4.3) has lost its physical meaning, and therefore 
there is little chance that one can obtain any physics out of numerical simulations of 
the lattice field theory. In this respect, it is, in a sense, not surprising that 
the calculation of Wilson's area law  in QED is incorrect \cite{asaga,fujita}. 

Since Wilson's calculation is presented analytically, it may be worth 
writing again the result of the Wilson loop calculation in QED 
$$ W \equiv 
 {\cal N} \prod_{m}\prod_\mu \int_{-\infty}^{\infty}dB_{m \mu} 
 \exp \left( i\sum_P   B_{n\mu}+{1\over 2g^2}
\sum_{n \mu \nu}e^{ if_{n \mu \nu}} \right).   \eqno{(4.8)} $$
In the strong coupling limit, one can evaluate eq.(4.8) analytically as 
$$  W \ \simeq (g^2)^{- \Delta S/a^2}  \eqno{(4.9)} $$
where $\Delta S$ should be an area encircled by the loop. This has the same behavior as 
that of eq.(4.6) since the lattice constant $a$ is equal to 
$ a= \Delta x =\Delta t $, that is, 
$$ W=f(g, \Delta S, \Delta x, \Delta t) .  \eqno{(4.10)}  $$
This amplitude $W$ has the dependence of the artificial parameters  
$\Delta t$ and $\Delta x $.  
This is completely different from eq.(4.7), and therefore one sees that the calculation 
of eq.(4.8) has no physical meaning, contrary to Feynman's treatment 
which has a right behavior as the function of the field parameters $q_{\bm{k},\lambda}$.  

\subsection{Physics of field quantization}

The quantization of fields is required from experiment as we stated in section 3. 
Yet, it is theoretically quite difficult to understand the basic 
physics of the field quantization. 
The fundamental step of the quantization is that the Hamiltonian one 
considers becomes an operator after the quantization. The reason why one considers 
the Hamiltonian is basically because it is a conserved quantity. In this respect, one 
cannot quantize the Hamiltonian density since it is not a conserved quantity yet. 
This is an important reason why one must quantize the field in terms of 
the creation and annihilation operator $ c_{\bm{k},\lambda}$ and  
$c^{\dagger}_{\bm{k},\lambda}$ in QED.  
In this respect, it is clear that the field quantization must be done 
in terms of $ c_{\bm{k},\lambda}$ and  $c^{\dagger}_{\bm{k},\lambda}$ with 
which the Hamiltonian in classical QED can be described. 

\subsection{No connection between fields and classical mechanics}

Here, we should make a comment on the discretized coordinates and fields. 
The discretized space is, of course, artificial, and there is no physics 
in the discretized fields and equations. In some textbook \cite{sakurai}, 
the field equation is derived from the picture that the field is constructed 
by the sum of springs in which the discretized coordinates of neighboring sites 
are connected by the spring. This picture can reproduce the field equation for 
a massless scalar field by adjusting some parameters, even though 
one started from a non-relativistic classical mechanics. However, 
this is obviously a wrong picture for a scalar field theory since the field equation 
has nothing to do with classical mechanics. It is somewhat unfortunate that it may 
have had some impact on the picture concerning lattice gauge field calculations 
as an excuse to make use of the discretized classical fields.  
As we saw in section 2, the path integral formulation has nothing 
to do with the dynamics of classical mechanics, and it is, of course, clear 
that the field theory path integral is never connected to any dynamics of 
the classical field theory. 

Concerning the scalar boson, one sees the difficulty that 
one cannot obtain the equation for a scalar field itself 
from the fundamental principle \cite{kkof} if it is an elementary field.  
In this respect, it should be a future problem to understand the problem of 
scalar fields more in depth \cite{fujita}.

\vspace{1cm}
\section{Partition function $Z$ in field theory  }

As we saw in section 3, Feynman's path integral formulation 
in terms of many dimensional integrations in parameter space  
is indeed a plausible method to quantize the fields in QED.  However, people commonly 
use the expression of the amplitude $Z$ as defined in eq.(4.1). 

\subsection{Partition function in QCD}
The ``new formulation" of the path integral in QCD was introduced by Faddeev and Popov 
who wrote the S-matrix elements as \cite{fp}
$$ \langle {\rm out}|{\rm in} \rangle \equiv Z= {\cal N} \int [{\cal D} A^a_\mu (x)] 
\exp \left[ i\int {\cal L}_{\rm QCD}  d^4x \right]\eqno{(5.1)} $$
where the definition of the path integral volume $ [{\cal D} A^a_\mu (x)] $ is just 
the same as the one explained in the previous section
$$ \int [{\cal D} A^a_\mu (x)] \equiv \lim_{n\rightarrow \infty} 
 \int_{-\infty}^\infty \cdots \int_{-\infty}^\infty \prod_{\mu,a} 
\prod_{(i,j,k,\ell)}^n dA^a_\mu (x_0^i,x_1^j,x_2^k,x_3^\ell). \eqno{(5.2)} $$
The Lagrangian density ${\cal L}_{\rm QCD} $ for QCD is given as 
$$ {\cal L}_{\rm QCD} =-{1\over 2} {\rm Tr} \left(G_{\mu \nu}G^{\mu \nu} 
\right) \eqno{(5.3)} $$
where $G_{\mu \nu}$ denotes the field strength in QCD 
$$ G_{\mu \nu}=\partial_\mu A_\nu -\partial_\nu A_\mu
+ig[A_\mu,A_\nu] .  \eqno{(5.4)} $$
However, the expression of the path integral in QCD in eq.(5.1) 
does not correspond to the field quantization. One can also understand why 
the path integral formulation of QCD cannot be done, in contrast to Feynman's 
integrations over the parameter space in QED. In QCD, the Hamiltonian for gluons contains 
the fourth power of $q_{\bm{k},\lambda}^a$, and therefore 
one cannot carry out the Gaussian integrations over the parameters   
$q_{\bm{k},\lambda}^a$ in QCD. As Feynman stated repeatedly in his original papers, 
the path integral formulation is closely connected to the Gaussian integration 
where the kinetic energy term in non-relativistic quantum mechanics is always 
described in terms of the quadratic term ${ \bm{p}^2\over 2m}$. 
This naturally leads to the conclusion that the path 
integral formulation cannot be properly constructed in QCD, and this is consistent 
with the recent investigation that QCD does not have a free Fock space due to its 
gauge non-invariance, and therefore one cannot carry out the field quantization of 
QCD Hamiltonian in perturbation theory \cite{fujita,fks1}. At the present stage, there 
is no solid method to calculate any physical observables in QCD since one may have 
to start from the total Hamiltonian of QCD which is, of course, gauge invariant but 
should be extremely difficult to treat.

\subsection{Fock space}

In quantum field theory, one must prepare Fock space since the Hamiltonian 
becomes an operator. The second quantized formulation is based on the creation 
and annihilation operators which act on the Fock space. In Feynman's 
path integral formulation, he prepared states which determine the number 
of photons in terms of $| q_{\bm{k},\lambda} \rangle $. Therefore, he started from 
the second quantized expression of the path integral formulation. 
However, Faddeev and Popov simply employed the same formula of the Lagrangian 
density, but integrated over the function  ${\cal D} A^a_\mu (x)$. 
This cannot specify any quantum numbers of the Fock space, and therefore 
the integration over the function  ${\cal D} A^a_\mu (x)$ does not 
correspond to the field quantization.

\vspace{1cm}
\section{Conclusions}

We have presented some basic problems in the path integral formulation in quantum 
mechanics as well as in quantum field theory. In the path integral of quantum mechanics, 
we clearly stated again that the path integral formulation has nothing to do 
with the classical paths in the corresponding classical mechanics. We must be 
careful for the superficial similarity between the path integral expression and 
the classical action. 

In the field theory case, we show that the standard procedure of the path integral method 
can not be related to any physical observables,  and therefore one can not make 
use of the path integral formulation in order to evaluate the energy spectrum in the field 
theory models. In order to clarify the problems in the path integral formulation, 
we have briefly reviewed Feynman's treatment of the path integral method in QED. This is 
physically plausible, and one can indeed calculate some of the S-matrix rules. In this 
case, however, Feynman employed the quantum mechanical treatment of the fermion 
path integral, and therefore the field quantization is only for the gauge field 
since the second quantization of the gauge field can be done in terms of the path 
integral over the parameter space in the vector potential.  In this respect, as Feynman 
himself repeatedly claimed in his original paper, the path integral method is, of course, 
not more than the quantum mechanics itself. In addition, there is no special advantage 
of employing the path integral formulation even in quantum mechanics. 
In the field theory case, the path integral over the parameter space is rather special 
for the QED case, and it is not at all clear whether the path integral method can be 
applied to other field theory models in the same way as Feynman's treatment. 
We believe that there is no chance to calculate any physical observables 
in QCD even in terms of Feynman's path integral formulation since the field energy 
of QCD should become the fourth power of the parameters $q_{\bm{k},\lambda}$, and 
this is impossible to carry out the path integrations even in the parameter space. 
However, this is consistent with the recent observation in QCD that the perturbation 
scheme is not well defined in QCD since the color charges of gluons are gauge 
dependent \cite{fujita,fks1}.

\vspace{0.7cm}

\section*{Acknowledgements}

The author is grateful to T. Asaga, A. Kusaka, S. Oshima, H. Takahashi and K. Tsuda 
for useful discussions and helpful comments. 


\end{document}